\documentclass[sigconf,screen]{acmart}
\AtBeginDocument{%
  }

\usepackage[normalem]{ulem}
\usepackage{enumitem}
\usepackage{xspace}
\usepackage{svg} 
\usepackage{multirow}
\usepackage{booktabs}
\usepackage{needspace}

\usepackage[most]{tcolorbox}

\newcommand{\method}{\textsc{DepTrans}\xspace}
\newcommand{\train}{\textsc{RAST-7B}\xspace}
\newcommand{\up}[1]{\textcolor{red}{\(\uparrow\)#1}}
\newcommand{\down}[1]{\textcolor{green}{\(\downarrow\)#1}}
\newcommand{\finding}[2]{
\begin{tcolorbox}[width=\linewidth,boxrule=0pt,top=1pt, bottom=1pt, left=1pt,right=1pt, colback=cyan!10,colframe=cyan!10, before skip=3pt,
after skip=3pt]
\textbf{Finding #1:} 
{#2}
\end{tcolorbox}
}

\newcommand{\cy}[1]{\textcolor{black}{#1}}

\newcommand{\gwj}[1]{\textcolor{black}{#1}}

\newcommand{\wcz}[1]{\textcolor{black}{#1}}

\acmYear{2026}\copyrightyear{2026}
\setcopyright{cc}
\setcctype[4.0]{by}
\acmConference[FSE Companion '26]{34th ACM International Conference on the Foundations of Software Engineering}{July 5--9, 2026}{Montreal, QC, Canada}
\acmBooktitle{34th ACM International Conference on the Foundations of Software Engineering (FSE Companion '26), July 5--9, 2026, Montreal, QC, Canada}
\acmDOI{10.1145/3803437.3805266}
\acmISBN{979-8-4007-2636-1/26/07}


\begin{document}

\title{Dependency-Guided Repository-Level C-to-Rust Translation with Reinforcement Alignment}



\author{Jia Feng}
\affiliation{%
  \institution{Harbin Institute of Technology}
  \city{Shenzhen}
  \country{China}
  }
\email{jiafeng@stu.hit.edu.cn}
\authornote{Equal contribution.}

\author{Wenjie Gan}
\affiliation{%
  \institution{Southeast University}
  \city{Nanjing}
  \country{China}
  }
\email{wenjiegan@seu.edu.cn}
\authornotemark[1]

\author{Cuiyun Gao}
\authornote{Corresponding author}
\affiliation{%
  \institution{Harbin Institute of Technology}
  \city{Shenzhen}
  \country{China}
  }
\email{gaocuiyun@hit.edu.cn}

\author{Chaozheng Wang}
\affiliation{%
  \institution{The Chinese University of Hong Kong}
  \city{Hong Kong}
  \country{China}
  }
\email{adf111178@gmail.com}

\author{Feng Luo}
\affiliation{%
  \institution{Harbin Institute of Technology}
  \city{Shenzhen}
  \country{China}
  }
\email{hitszluofeng@foxmail.com}

\author{Xin Xia}
\affiliation{%
  \institution{Zhejiang University}
  \city{Hangzhou}
  \country{China}
  }
\email{xin.xia@acm.org}

\author{Ge Li}
\affiliation{%
  \institution{Peking University}
  \city{Beijing}
  \country{China}
  }
\email{lige@pku.edu.cn}

\author{Kui Liu}
\affiliation{%
  \institution{Huawei}
  \city{Shenzhen}
  \country{China}
  }
\email{kui.liu@huawei.com}

\begin{abstract}
Automating C-to-Rust migration is critical for enhancing software security without compromising performance. Traditional rule-based methods often struggle with diverse C idioms, yielding unidiomatic and rigid Rust translations. In contrast, Large Language Models (LLMs), enriched by massive code corpora, offer a superior alternative due to their cross-language generalization, enabling the generation of idiomatic and maintainable Rust code.
However, applying LLMs to C-to-Rust migration still poses several challenges. First, existing LLM-based approaches fail to adequately handle cross-file references. \cy{They either overlook the dependencies or incorporate entire files as context, thereby hindering the model's ability to accurately capture dependency information for translation. Second, the complex dependencies and structured inputs/outputs make it difficult to validate the syntactic correctness and functional equivalence of repository-level translations.} 
\cy{Furthermore, the scarcity of large-scale C-Rust parallel data constrains the inherent generation capabilities of LLMs, resulting in limited performance.}
To address these challenges, we propose \method, which synergizes model-level capability enhancement with a structured inference framework. \method first employs \cy{the} \textit{Reinforcement-Aligned Syntax Training} \cy{module} to bolster the model's intrinsic \cy{generation capabilities}
through
multi-task fine-tuning and
feedback-driven reinforcement learning. Building upon this enhanced foundation \cy{model}, the \textit{Dependency-Guided Iterative Refinement} module \cy{identifies the fine-grained cross-file dependencies as enriched context for the initial translation, and further iteratively refines the generated Rust code to ensure the syntactic correctness and functional equivalence.}
To facilitate \cy{the model training and repository-level evaluation},
we construct a large-scale corpus of 85k \cy{training} samples
and a specialized benchmark of 145 repository-level instances.
Experimental results show that \method achieves a compilation success rate of 60.7\% and a computational accuracy of 43.5\%, outperforming the strongest baseline by 22.8\% and 17.3\%, respectively.
Moreover, \method successfully builds 7 out of 15 industrial-scale C projects from Huawei’s internal codebase, demonstrating its 
potential
in enterprise-level C-to-Rust migration.

\end{abstract}


\maketitle
\section{Introduction}

System-level development demands a rigorous balance between performance and memory safety~\cite{DBLP:conf/uss/CriswellGA09}. While C provides the necessary low-level control, its inherent lack of memory safety remains a primary source of critical vulnerabilities~\cite{msrc2019proactive}. Rust has emerged as a compelling alternative, offering C-equivalent performance while enforcing memory safety via its static ownership system~\cite{rust_wiki,lesinski_speed}. Consequently, migrating legacy C codebases to Rust is now a strategic priority for enhancing software security~\cite{darpa_tractor,DBLP:journals/ieeesp/Larsen24}. However, the manual migration process is prohibitively labor-intensive and error-prone, necessitating the development of robust, automated C-to-Rust translation solutions~\cite{DBLP:conf/ndss/LiWLSK25}.

In recent years, 
tools like C2Rust~\cite{c2rust_website} have automated C-to-Rust migration by mapping C syntax to Rust constructs.
However, 
these rule-based methods
often produce Rust code that retains C-style design~\cite{hong2025forcratautomaticioapi}, such as raw pointers and unsafe blocks ~\cite{DBLP:conf/kbse/HongR24,DBLP:journals/corr/abs-2505-04852}. Although the output is typically compilable, it is frequently unidiomatic~\cite{DBLP:journals/corr/abs-2412-15042}, requiring developers to invest manual efforts in post-translation refactoring. 
LLMs have recently shown strong performance on code-related tasks
~\cite{DBLP:conf/acl/HuangL0D24,10.1145/3728933,DBLP:journals/tmlr/ZhangCLLG0L024,10.1145/3729341}. 
Previous studies \cite{DBLP:journals/corr/abs-2503-12511,luo2025integrating,VERT} have shown that compared to rule-based tools such as C2Rust, LLMs offer clear advantages in producing Rust code that better aligns with the idiomatic practices of the language, significantly reducing the manual refactoring~\cite{DBLP:journals/corr/abs-2501-14257,DBLP:journals/corr/abs-2503-12511,rustmap}.
Despite the effectiveness of LLMs, existing LLM-based approaches also face critical challenges. 
First, LLMs frequently struggle with the fine-grained dependencies of large-scale repositories, leading to fragmented and inconsistent translations~\cite{yun2024project,feng2024complexcodeeval,li2025aixcoder}. Simply translating functions in isolation or
cramming entire files into the prompt 
typically yields suboptimal results,
as the model fails to maintain a coherent global view across file boundaries, resulting in broken references and inconsistent program logic~\cite{pan2024lost,wang2025apirat}. 
Second, the complex dependencies and structured inputs/outputs of repository-level software make it exceptionally difficult to validate syntactic correctness and functional equivalence. The reliance on intricate data structures defined across multiple modules hinders the generation of standalone test cases, thereby depriving the model of the high-quality diagnostic feedback---such as compiler diagnostics or execution traces---essential for driving effective iterative self-refinement~\cite{sim2025large,yang2025constructing,ferrara2024challenges}.
Finally, 
the scarcity of parallel C-Rust repositories in training corpora hinders models from learning 
the complex mappings required for cross-file coherence. Without sufficient exposure to diverse repository-level patterns, LLMs frequently generate code with unresolved references or fragmented logic. This lack of architectural awareness leads to significantly degraded compilation success rates and functional accuracy in large-scale migrations~\cite{li2025aixcoder,assogbaevaluating,liu2025comprehensive}.
To address these challenges, we propose \method, a dependency-guided approach for repository-level C-to-Rust translation. The core of \method lies in its synergy between model-level capability enhancement and a structured inference framework.
First, we introduce \textit{Reinforcement-Aligned Syntax Training} (RAST) to bolster the model's intrinsic dependency awareness and task-specific adaptation. This two-stage scheme employs multi-task fine-tuning to capture distant structural correlations, followed by compiler-feedback reinforcement learning to further enforce syntactic correctness and functional equivalence. By optimizing the model's fundamental translation and repair capabilities, RAST provides a robust foundation for handling complex repository structures.
Second, building upon the enhanced model, we develop a \textit{Dependency-Guided Iterative Refinement} (DGIR) framework to orchestrate the translation process. This involves a \textit{Cross-Language Dependency Alignment} strategy that decomposes C projects into manageable units and performs incremental, bottom-up translation following topological orders. By mapping extracted C dependencies to a semantically aligned Rust dependency pool, the framework provides precise, fine-grained context. Furthermore, a \textit{Consistency-Guided Refinement} strategy iteratively optimizes the translation by integrating compiler diagnostics with LLM-based self-consistency checks, ensuring the output is both syntactically valid and functionally aligned.

To facilitate both training and evaluation, we construct a large-scale corpus of 85k instances, comprising 82.5k instances for multi-task fine-tuning and 2.5k high-quality instances for reinforcement learning. For evaluation, we curate a specialized benchmark of 145 repository-level instances with complex cross-file dependencies to rigorously assess contextual reasoning.
Experimental results demonstrate that even without RAST's task-specific adaptation, DGIR alone enables Qwen2.5-Coder-32B to achieve a 57.2\% compilation success rate and 41.4\% computational accuracy, outperforming the strongest baseline by 19.3 and 15.2 percentage points, respectively. Remarkably, after applying our RAST, a smaller Qwen2.5-Coder-7B reaches 60.7\% compilation success and 43.5\% accuracy, surpassing the untrained 32B model. 
Finally, validation on 15 industrial-scale projects from Huawei’s database system shows that our model successfully migrates 7 projects to a buildable state,
demonstrating the potential of \method for enterprise-level C-to-Rust migration.

The main contributions of this paper are as follows.
\begin{enumerate}[label=\arabic{*}), topsep=0pt]
\item We introduce a two-stage training paradigm combining multi-task fine-tuning with compiler-feedback reinforcement learning, enhancing the model's cross-file dependency awareness and task-specific adaptation.

\item We propose a dependency-guided framework that integrates fine-grained dependency with consistency-driven iterative refinement to ensure syntactic and functional correctness in repository-level translation.


\item We construct a large-scale corpus of 85k training samples and a dependency-intensive benchmark of 145 repository-level samples, providing a foundation for both training and evaluating C-to-Rust translation.

\item Extensive experiments demonstrate our approach's effectiveness in dependency comprehension and its practical potential for enterprise-level migration.






\end{enumerate}

Our source code and experimental data are publicly available at 
https://github.com/2726fj/C2RustRep.

\section{Methodology}

\begin{figure*}[htbp]
    \centering
    \includegraphics[width=0.9\textwidth]{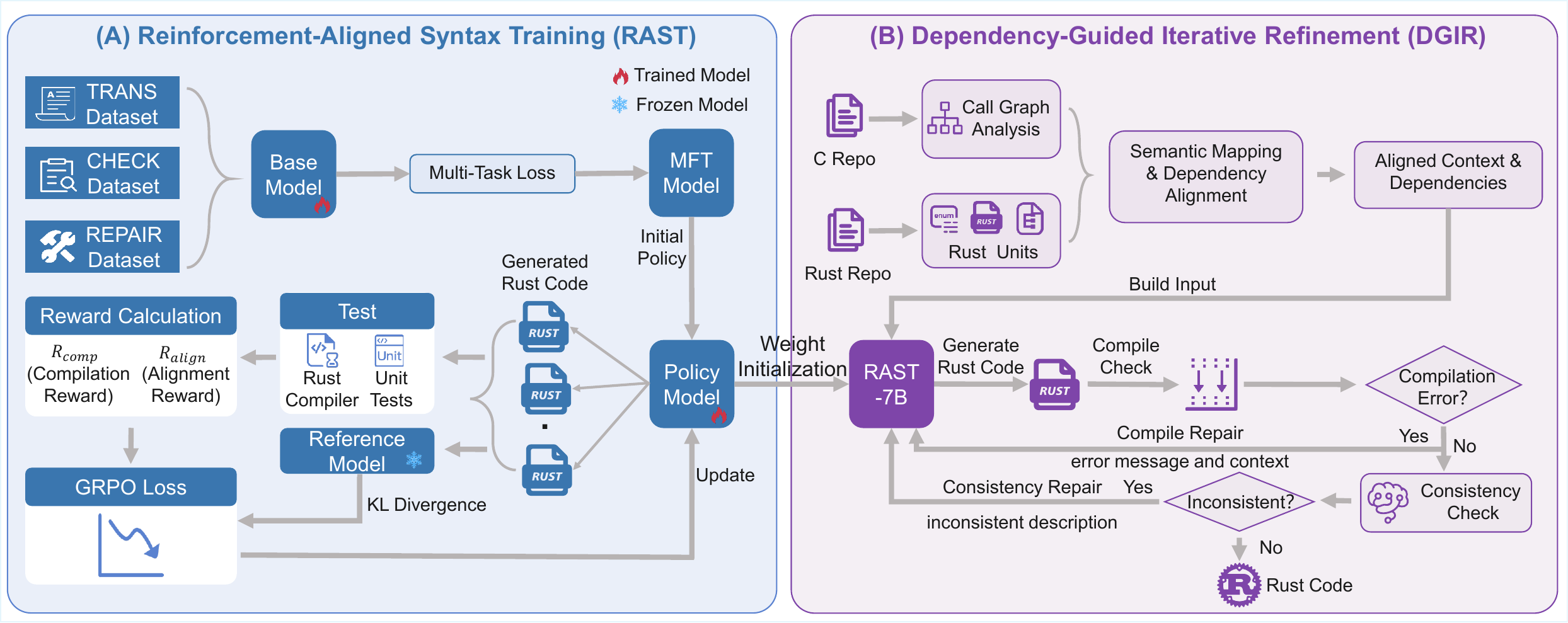}
    \caption{Overview of \method.}
    \Description{Overview of the proposed framework, showing the dependency-guided C-to-Rust translation process with model training, dependency alignment, and iterative refinement modules.}
    \label{fig:framework}
\end{figure*}


This section details the architecture of \method, which integrates model-level training with a structured inference framework. We first present the \textit{Reinforcement-Aligned Syntax Training} in Section~\ref{sec:training_paradigm}, designed to bolster the model's intrinsic dependency awareness and self-correction capabilities. Building upon this foundation, Section~\ref{sec:infer_framework} introduces the \textit{Dependency-Guided Iterative Refinement} framework, which serves as the inference engine to orchestrate repository-level translation and compiler-driven refinement. 

\subsection{Reinforcement-Aligned Syntax Training}
\label{sec:training_paradigm}

While the inference framework provides structural guidance, the model's intrinsic ability to handle repository-level nuances often remains a bottleneck. LLMs frequently struggle with cross-file structural dependencies and lack a specialized ``translation-detection-repair'' synergy. To bridge this gap, we propose \textit{Reinforcement-Aligned Syntax Training} (RAST), a two-stage paradigm designed to bolster the model's fundamental awareness of global repository logic and its adaptive capacity for the migration task, as shown in \autoref{fig:framework} (part A).

\subsubsection{Multi-Task Fine-Tuning}
While the inference framework provides structural guidance, the model's intrinsic ability to handle repository-level nuances often remains a bottleneck. LLMs frequently struggle with cross-file structural dependencies and lack a specialized synergy between translation, detection, and repair. To bridge this gap, we formulate the initial training stage as a multi-task learning problem. This approach forces the model to internalize the relationship between C logic and Rust syntax while developing an acute sensitivity to migration pitfalls.

Specifically, the model simultaneously learns three interlinked tasks: (1) \textit{C-to-Rust Translation}, where the model generates semantically equivalent Rust code given a C source and its dependency context; (2) \textit{Syntax Error Detection}, where the model assesses Rust code integrity and provides diagnostic feedback; and (3) \textit{Code Repair}, where the model corrects erroneous Rust snippets based on error descriptions. To enable task-specific adaptation, each training sample is prepended with a distinct \texttt{[TaskTag]}. We employ a unified autoregressive language modeling objective:
\begin{equation}
\mathcal{L}_{\text{MTL}} = -\sum_{i=1}^{T} \log P_\theta(y_i \mid y_{<i}, x, \texttt{[TaskTag]})
\label{eq:mtl}
\end{equation}
where $x$ is the input prompt, $y$ is the target output sequence, and $\theta$ is the model parameters. 
By jointly optimizing these objectives, the model acquires the requisite ``inner logic'' to effectively power the \textit{Dependency-Guided Iterative Refinement} framework, as detailed in Section~\ref{sec:infer_framework}. This synergistic training transforms the model from a passive generator into an active, self-correcting agent capable of autonomous architectural reasoning.

\subsubsection{Compiler-Feedback Reinforcement Learning}

While multi-task fine-tuning establishes a foundational mapping, the maximum likelihood objective often prioritizes token-level probability over the rigorous, binary requirements of a compiler. To bridge the gap between "plausible" and "executable" code, we introduce a reinforcement learning stage using Group Relative Policy Optimization (GRPO) \cite{guo2025deepseek}. This stage aligns the model with the strict type system of Rust by treating the compiler and test suite as an interactive environment for policy optimization.

\paragraph{Optimization Objective.}  

For each question $q$, GRPO samples a group of outputs $\{o_1, o_2, \dots, o_G\}$ from the old policy $\pi_{\theta_{old}}$ and optimizes the policy model $\pi_\theta$ by maximizing the following objective:

\begin{equation}
\begin{aligned}
\mathcal{J}_{GRPO}(\theta) = \mathbb{E} & \Big[ q \sim P(Q), \{o_i\}_{i=1}^G \sim \pi_{\theta_{old}}(O|q) \Big] \\
\frac{1}{G} \sum_{i=1}^G \Big( \min & \left( \frac{\pi_\theta(o_i|q)}{\pi_{\theta_{old}}(o_i|q)} A_i, \text{clip} \left( \frac{\pi_\theta(o_i|q)}{\pi_{\theta_{old}}(o_i|q)}, 1-\varepsilon, 1+\varepsilon \right) A_i \right) \\
& - \beta \mathbb{D}_{KL}(\pi_\theta || \pi_{ref}) \Big)
\end{aligned}
\label{eq:grpo_objective}
\end{equation}

where the advantage $A_i$ is computed by normalizing the rewards within each group to emphasize relative quality, thereby eliminating the need for a separate value network:
\begin{equation}
A_i = \frac{r_i - \text{mean}(\{r_1, r_2, \dots, r_G\})}{\text{std}(\{r_1, r_2, \dots, r_G\})}
\label{eq:advantage}
\end{equation}

The KL divergence term is efficiently estimated as:
\begin{equation}
\mathbb{D}_{KL}(\pi_\theta || \pi_{ref}) = \frac{\pi_{ref}(o_i|q)}{\pi_\theta(o_i|q)} - \log \frac{\pi_{ref}(o_i|q)}{\pi_\theta(o_i|q)} - 1
\label{eq:kl_div}
\end{equation}
where $\pi_{ref}$ is the SFT-tuned reference policy. This objective ensures training stability by preventing the policy from deviating excessively from the reference.

\paragraph{Reward Design.}  

To reinforce both syntactic and semantic integrity, we design a \textbf{hybrid reward function} $R(a) = \alpha R_{\text{comp}} + \beta R_{\text{align}}$.
\begin{itemize}
    \item \textbf{Syntactic Validity ($R_{\text{comp}}$):} To penalize invalid syntax, we define $R_{\text{comp}} = \frac{1}{1 + N_{\text{err}}}$, where $N_{\text{err}}$ is the count of diagnostic errors from the Rust compiler. This forces the model to prioritize compilability.
    \item \textbf{Functional Alignment ($R_{\text{align}}$):} For function-level samples with test cases, $R_{\text{align}}$ is the pass rate of unit tests. For repository-level samples where execution is infeasible, $R_{\text{align}}$ is the \textit{CodeBLEU} score against the reference, which ensures structural and data-flow consistency in migration scenarios.
\end{itemize}


Guided by the need for computational efficiency, we implement the RAST paradigm on Qwen2.5-Coder-7B~\cite{hui2024qwen2}. By assigning balanced reward weights ($\alpha = 1, \beta = 1$), we prioritize syntactic and functional integrity equally, demonstrating that RAST can elicit expert-level capabilities even from smaller-scale models. We denote this optimized version as \textbf{\train}, an efficient engine for repository-level translation. The sensitivity of reward weights is further discussed in Section~\ref{sec:ablation_train}.

\subsection{Dependency-Guided Iterative Refinement}
\label{sec:infer_framework}


To address the complexities of repository-scale migration, we design the \textit{Dependency-Guided Iterative Refinement} (DGIR) framework as the core inference engine of \method. As illustrated in~\autoref{fig:framework} (part B), the framework operates through two synergistic modules: 
First, the \textit{Cross-Language Dependency Alignment} Module ensures structural consistency by performing an incremental, bottom-up translation following the topological order of dependencies. It extracts precise context from both C call graphs and a categorized Rust dependency pool to guide the LLM's initial generation. 
Second, the \textit{Consistency-Guided Translation Refinement} Module capitalizes on the LLM's self-refinement potential. It utilizes the aligned dependencies alongside compiler diagnostics and self-consistency checks to iteratively repair errors, ensuring that the final output achieves high syntactic validity even in the absence of extensive test suites.
Each component is described in detail in the following sections.

\subsubsection{Cross-Language Dependency Alignment}
\label{subsec:Dependency_Modeling}

Repository-level translation is hindered by the fragmentation of semantic context across file boundaries. Without explicit dependency modeling, LLMs often generate code that references undefined symbols or violates target-language module structures. To mitigate this, we propose a \textit{Cross-Language Dependency Alignment} module that bridges the semantic gap between the source C repository and the target Rust environment through three steps.

\textbf{1. Dependency Graph Construction in C.}
To determine the optimal translation sequence and capture the logical scope of each function, we first model the structural dependencies of the C project. Using Tree-Sitter \cite{tree-sitter}, we construct a \textit{global call graph} by traversing all source files and mapping inter-function relationships. The motivation behind this graph is twofold: it allows us to identify the \textit{topological order} for incremental bottom-up translation, and it enables the extraction of a comprehensive \textit{dependency set} for each function. This set includes not only invoked functions but also critical global context such as headers, global variables, structs, and macro definitions.

\textbf{2. Categorized Dependency Pooling in Rust.}
While C dependencies provide the "requirement," the target Rust environment provides the "available resources." To facilitate precise mapping, we build a structured \textit{Rust-side dependency pool} by extracting and categorizing elements into ten syntactic types, including \texttt{struct}, \texttt{enum}, \texttt{function}, and \texttt{trait}. Critically, for types such as \texttt{struct} and \texttt{trait}, we explicitly link their associated \texttt{impl} blocks. This categorization is essential because Rust's modularity and method implementation patterns differ significantly from C's procedural structure, requiring a more granular organization of target-side assets.

\textbf{3. Semantic Alignment and Augmentation.} 
The final step is to resolve the cross-language mapping between C requirements and Rust implementations. We employ \textit{BGE-M3}~\cite{bge-m3} to vectorize all Rust structures, leveraging its superior cross-lingual semantic representation to overcome syntactic divergence. For each C dependency, we perform a cosine similarity-based retrieval from the Rust pool to find its semantically equivalent counterpart. To ensure \textit{semantic completeness}, if a retrieved Rust function resides within an \texttt{impl} block, we automatically augment the context with the corresponding parent structure. This look-ahead mechanism prevents the generation of "orphaned" methods and ensures that the LLM receives a logically complete context for translation.

\subsubsection{Consistency-Guided Translation Refinement}
Direct LLM translation at the repository level often produces "hallucinated" syntax or logic shifts that violate cross-file invariants. To bridge this gap, we propose a two-stage refinement module designed to maximize the LLM's self-refinement potential. By providing structured contextual grounding and multi-dimensional feedback, the module empowers the model to iteratively identify and rectify its own translation errors.

\textbf{1. Dependency-Aware Context Construction.}
To provide the necessary "knowledge base" for effective self-refinement, we transform aligned dependencies into a granularity-adaptive prompt. Following the \textit{topological order} from Section~\ref{subsec:Dependency_Modeling}, we process functional units incrementally. To avoid distracting the model with excessive noise, large structural dependencies (e.g., extensive structs) are abstracted into concise docstrings, while smaller, pivotal code segments are provided verbatim. Furthermore, the LLM is prompted to generate a cross-language bridge docstring that maps C logic to Rust-idiomatic intent. This hybrid context ensures the model has a clear semantic blueprint to guide its initial generation and subsequent refinement iterations.

\textbf{2. Iterative Repair via Multi-Dimensional Feedback.}
Recognizing that LLMs perform best when given explicit "hints" about their mistakes, we implement a feedback-driven repair loop that acts as the primary engine for self-refinement.
\begin{itemize}
    \item \textit{Diagnostic-Driven Repair:} Upon compilation, syntax or type errors are fed back to the LLM. This provides a precise corrective signal, allowing the model to refine the Rust code until it satisfies the rigorous constraints of the Rust compiler.
    \item \textit{Consistency-Guided Verification:} In the absence of executable test suites, we utilize semantic consistency checks \cite{wang2025can} to provide a high-level refinement signal. By acting as a dual-language auditor, the LLM compares the C source with its Rust translation to uncover logical discrepancies. These mismatches serve as a heuristic guide for self-correction, enabling the model to iteratively align the functional behavior of the two languages without requiring external test oracles.
\end{itemize}

\section{\cy{Training and Evaluation Data}
Construction}
\label{sec:dataset}
To address \cy{the} data scarcity \cy{issue} and bolster \cy{the inherent generation capabilities of LLMs},
we curated a multi-granularity dataset for C-to-Rust migration\cy{, including the repository-level and function-level corpora.}

\subsection{Repository-Level Dataset Construction}
\label{sec:repo_level}
To capture authentic cross-file dependencies, we curate a high-quality repository-level dataset through a three-phase pipeline, focusing on real-world migration logic.

\textbf{Phase 1: Project Selection and Structural Parsing.} We identify candidate repository pairs by searching GitHub for keywords such as ``migration,'' ``rewrite,'' and ``C to Rust.'' We select projects where both versions are maintained by the same organization to ensure functional alignment. Following common practice~\cite{li2025aixcoder}, we filter for projects with $>10$ stars and verified buildability. Beyond file parsing, we use Tree-Sitter~\cite{tree-sitter} and call graph analysis to retain only functions with active cross-function dependencies, ensuring the dataset captures true repository-level complexity.

\textbf{Phase 2: Hybrid Semantic Alignment.} To identify equivalent functions, we employ a three-step filtering process: (1) \textit{Coarse Retrieval}: We utilize BGE-M3~\cite{bge-m3}, a widely-used model for cross-lingual code retrieval, to efficiently retrieve top candidates via cosine similarity. (2) \textit{Model-based Re-ranking}: We leverage Qwen2.5-Coder-32B to perform semantic verification, exploiting its reasoning capabilities to filter out false positives. (3) \textit{Manual Audit}: The first and second authors conduct a final review of the aligned pairs within the full repository context to resolve any subtle discrepancies and ensure high-fidelity ground truth.

\textbf{Phase 3: Test-based Coverage Identification.} To determine which functions are verifiable via execution, we employ a straightforward function-level deletion strategy. For each Rust function, we temporarily remove its body and execute \texttt{cargo test}. If this leads to a test failure, we interpret it as a definitive signal that the function's logic is covered by the existing test suite, marking it as a ``test-verifiable'' sample.

\subsection{Function-Level Dataset Construction}
\label{sec:func_level}
To overcome the scarcity of parallel repositories, we construct a large-scale synthetic dataset to bolster the model's fundamental translation and self-correction capabilities.

\textbf{Phase 1: Seed Selection and Diversity Filtering.} We utilize \textit{xCodeEval}~\cite{xcodeeval} as our primary source, as it provides a vast collection of C functions with associated test cases. We filter for functions with token counts between 64 and 2,048 to exclude trivial boilerplate code while ensuring samples fit within standard LLM context windows. Inspired by prior work on data deduplication~\cite{abbas2023semdedup}, we ensure semantic diversity by generating embeddings with BGE-M3~\cite{bge-m3} and applying $K$-means clustering. This process allows us to sample representative functions across diverse C idioms, preventing the model from over-fitting on repetitive logic patterns.

\textbf{Phase 2: Compiler-Guided Synthesis.} To obtain high-quality C-Rust pairs, we employ an automated translation-and-refinement pipeline. Each C function is initially translated by an LLM, followed by an iterative optimization process guided strictly by Rust compiler diagnostics. This strategy ensures that the ground-truth authority rests on the rigorous compiler rather than the generative model, resulting in syntactically valid and idiomatic Rust counterparts.

\textbf{Phase 3: Multi-Task Label Generation.} Based on the synthesis trajectories, we derive labels for three synergistic tasks: (1) \textit{Translation}, using successfully compiled C-Rust pairs as ground truth; (2) \textit{Detection}, using both successful and failed attempts to train the model on error identification; and (3) \textit{Repair}, using the logs of the refinement process to provide realistic error-correction trajectories for supervised fine-tuning.

\subsection{Dataset Statistics}
\begin{table}[t]
\centering
\caption{
Statistical summary of the datasets across different
stages. "Max", "Min", and "Average" \cy{refer to the maximum, minimum and average}
token counts per sample, respectively.
}
\vspace{-0.3cm}
\renewcommand{\arraystretch}{0.8}
\begin{tabular}{@{}cccccc@{}}
\toprule
\multicolumn{1}{c|}{Type}                        & \multicolumn{1}{c|}{Dataset Name} & \#Samples & Max   & Min & Average \\ \midrule
\multicolumn{6}{c}{Multi-task Fine-tuning}                                                                               \\ \midrule
\multicolumn{1}{c|}{\multirow{3}{*}{Function}} & \multicolumn{1}{c|}{SyntaxCheck}  & 32,066    & 1,030 & 66  & 182     \\
\multicolumn{1}{c|}{}                            & \multicolumn{1}{c|}{CodeTrans}    & 32,066    & 1,030 & 66  & 182     \\
\multicolumn{1}{c|}{}                            & \multicolumn{1}{c|}{CodeFix}      & 18,743    & 1,033 & 80  & 242     \\ \midrule
\multicolumn{6}{c}{Reinforcement Learning}                                                                               \\ \midrule
\multicolumn{1}{c|}{Function}                    & \multicolumn{1}{c|}{FucnTrans}    & 2,370     & 1,921 & 30  & 305     \\
\multicolumn{1}{c|}{Repository}                  & \multicolumn{1}{c|}{RepoTrans}    & 310       & 1,717 & 16  & 257     \\ \midrule
\multicolumn{6}{c}{Evaluation}                                                                                           \\ \midrule
\multicolumn{1}{c|}{\multirow{2}{*}{Repository}} & \multicolumn{1}{c|}{DCBench}      & 125       & 4,769 & 31  & 316     \\
\multicolumn{1}{c|}{}                            & \multicolumn{1}{c|}{IMCBench}     & 20        & 453   & 31  & 157     \\ \bottomrule
\end{tabular}
\renewcommand{\arraystretch}{1}
\vspace{-0.3cm}
\label{tab:dataset_stats}
\end{table}

Following the construction pipelines detailed in Section~\ref{sec:repo_level} and Section~\ref{sec:func_level}, we curate a multi-stage dataset to support RAST and rigorous evaluation. \autoref{tab:dataset_stats} provides a comprehensive breakdown of the sample counts and token distributions.

\textbf{Multi-task Fine-tuning Dataset.} To equip the model with fundamental translation and self-repair capabilities, we construct three specialized tasks under the function-level category: (1) \textit{CodeTrans} for C-to-Rust translation, (2) \textit{SyntaxCheck} for error detection and analysis, and (3) \textit{CodeFix} for iterative code repair. These tasks, totaling over 82k samples, provide a balanced coverage of syntactic and functional patterns required for cross-language migration.

\textbf{Reinforcement Learning Dataset.} To further align the model with compilation and functional requirements, we utilize a hybrid RL dataset. This includes 2,370 function-level samples (Function) and 310 repository-level samples (Repository). The latter specifically incorporates cross-file dependencies to bolster the model's ability to navigate architectural constraints during the alignment process.

\textbf{Evaluation Benchmarks.} For a robust assessment of repository-level migration, we establish two benchmarks under the Repository category derived from parallel C/Rust projects. \textit{DCBench}, sourced from the \textit{deltachat-core} repository, comprises 125 samples with an average of 316 tokens. Its complex dependencies and 175-file scale provide a highly challenging testbed for contextual reasoning. In contrast, \textit{IMCBench}, derived from \textit{incubator-milagro-crypto}, contains 20 samples from a 121-file codebase, serving as an additional setting to evaluate basic dependency handling.

\section{Experimental setup}

\subsection{Selected LCMs}

We evaluate our framework across two state-of-the-art Large Code Model families. 
\textit{DeepSeek-Coder (DSC)}~\cite{guo2024deepseek} represents a LLaMA-based decoder-only architecture, for which we select the 6.7B and 33B instruction-tuned variants. 
\textit{Qwen2.5-Coder (QWC)}~\cite{hui2024qwen2} provides extensive long-context support (128K tokens), and we utilize its 7B, 14B, and 32B versions. 
This diverse configuration allows for a comprehensive analysis of how model scale and architectural capabilities influence syntactic validity and semantic fidelity in complex C-to-Rust migration.

\subsection{Research Questions}




In this work, we conduct comprehensive experiments to answer the following research questions:


\begin{itemize}
    \item \textbf{RQ1}: How does \method perform in C-to-Rust translation with intricate repository-level dependencies?
    \item \textbf{RQ2}: How do DGIR components and \train reward configurations affect migration quality?
    \item \textbf{RQ3}: Can \method handle real-world, large-scale industrial projects with stringent reliability requirements?
\end{itemize}

To address \textbf{RQ1}, we evaluate both our DGIR and \train on DCBench and IMCBench, which feature varying degrees of rich, cross-file dependencies ranging from standalone modules to intricate architectures.
To answer \textbf{RQ2}, we perform a series of ablation studies, systematically decoupling the DGIR's components and varying the RAST reward signals to quantify their respective impacts on syntactic and semantic performance. 
For \textbf{RQ3}, we assess the practical utility of \method on \textsc{HWBench}, a benchmark comprising 15 industrial C projects from Huawei's database systems. With codebases ranging from 1,000 to 3,000 lines.

\subsection{Compared Methods}
\label{sec:baselines}


\subsubsection{Baselines in RQ1.}
We compare DGIR with three baselines to evaluate the effectiveness of context capture and utilization: 
(1) \emph{Base}, which translates C to Rust without auxiliary context; 
(2) \emph{RAG}, which uses retrieval-augmented generation via fine-tuned BGE-M3 ($k=1,3,5$); 
and (3) \emph{File Context}, which provides the complete C/Rust file pairs (excluding target functions) as input. 
By focusing on these baselines, we isolate the impact of structured dependency guidance from raw or retrieved context. This setup is chosen because existing C-to-Rust tools are primarily designed for function-level tasks and lack the architectural capacity to handle the repository-scale dependencies our framework addresses.
To evaluate model-level enhancements, we compare \train with baseline models of varying scales. All candidates are integrated into the same DGIR framework to isolate the impact of our RAST paradigm from inference-level optimizations, thereby ensuring a controlled and fair assessment of the model's intrinsic migration capabilities.

\subsubsection{Ablation Studies for RQ2.}
To quantify the individual contribution of each component within DGIR, we evaluate four variants: 
(1) \emph{Plain Deps}, which provides raw dependencies without fine-grained structure rewriting or import recovery; 
(2) \emph{w/o Compile}, which disables compiler-guided repair; 
(3) \emph{w/o Consistency}, which removes semantic consistency checking; 
and (4) \emph{w/o Repair}, which disables all iterative refinement. 
This systematic ablation allows us to disentangle the performance gains attributed to structured dependency modeling versus iterative self-correction.
For \train, we investigate the impact of different reward weighting strategies by varying the ratio $\alpha:\beta$: 
\emph{Comp-Only (1:0)}, focusing exclusively on syntactic validity; 
\emph{Align-Only (0:1)}, prioritizing functional consistency; 
\emph{Balanced (1:1)}, our default setting treating both signals equally; 
and \emph{Align-Heavy (1:2)}, which emphasizes semantic correctness.

\subsubsection{Evaluation for RQ3.}
To evaluate practical effectiveness in real-world scenarios, we compare \method against two project-level strategies: 
(1) \emph{Base}, which translates all C files simultaneously following previous work's setting ~\cite{CRUST-Bench}; 
and (2) \emph{Sactor}~\cite{DBLP:journals/corr/abs-2503-12511}, a two-stage pipeline combining C2Rust~\cite{c2rust_website} static translation with LLM-based refinement. 
This comparison demonstrates that \method's integrated dependency modeling handles complex module relationships more effectively than naive global prompting or rigid multi-stage pipelines.


\subsection{Performance Metrics}
We adopt three metrics: 
(1) \textbf{Compilation Success Rate (CSR)}~\cite{luo2025integrating}, the proportion of compilable translations; 
(2) \textbf{Computational Accuracy (CA)}~\cite{yang2024,luo2025integrating}, functional equivalence with reference outputs; 
and (3) \textbf{CodeBLEU}~\cite{ren2020codebleu}, structural and lexical similarity to reference Rust code. 
For \textsc{HWBench}, we report \textbf{Num Build}~\cite{CRUST-Bench} (projects compiled), since these projects lack Rust test cases.

\subsection{Implementation Details}

Experiments are run on four NVIDIA A100 (40GB) GPUs. 
For multi-task fine-tuning, we use a learning rate of \texttt{1e-4} with cosine decay; for reinforcement alignment, \texttt{5e-5}. 
The per-device batch size is 2 with gradient accumulation of 8, and models are trained for 3 epochs. 
We apply sequence packing, gradient checkpointing, and LoRA ($r=16$, scaling=32, dropout=0.1) for efficiency, and adopt ZeRO-3 in DeepSpeed for memory optimization. 
Inference uses decoding with \texttt{temperature=0}, \texttt{top\_p=1}, and max 4096 tokens.

\begin{table*}[t]
\centering
\caption{
Evaluation results of DGIR compared with baselines on the DCBench and IMCBench datasets. Unless otherwise specified, bold indicates the best performance under each metric in all subsequent tables. The \up{}/\down{} values indicate DGIR's gain or loss over the best-performing baseline in computational accuracy (CA) and compilation success rate (CSR).
}
\vspace{-0.3cm}
\renewcommand{\arraystretch}{0.8}
\begin{tabular}{@{}c|c|cc|cc|cc|cc|cc@{}}
\toprule
\multirow{2}{*}{Dataset}  & \multirow{2}{*}{Method} & \multicolumn{2}{c|}{DSC-6.7B} & \multicolumn{2}{c|}{DSC-33B} & \multicolumn{2}{c|}{QWC-7B} & \multicolumn{2}{c|}{QWC-14B} & \multicolumn{2}{c}{QWC-32B} \\ \cmidrule(l){3-12} 
                          &                         & CSR           & CA            & CSR           & CA           & CSR          & CA           & CSR           & CA            & CSR           & CA            \\ \midrule
\multirow{7}{*}{DCBench}  & Base                    & 4.0           & 2.4           & 8.0           & 4.8          & 8.8          & 4.8          & 9.6           & 5.6           & 8.8           & 4.8           \\
                          & RAG (k=1)               & 6.4           & 3.2           & 8.8           & 6.4          & 8.0          & 3.2          & 12.0          & 5.6           & 10.4          & 4.8           \\
                          & RAG (k=3)               & 10.4          & 4.8           & 8.0           & 5.6          & 8.8          & 3.2          & 12.0          & 5.6           & 10.4          & 4.8           \\
                          & RAG (k=5)               & 7.2           & 3.2           & 7.2           & 4.0          & 8.0          & 4.8          & 15.2          & 8.0           & 12.0          & 6.4           \\
                          & File Context            & 21.6          & 15.2          & 18.4          & 16.0         & 20.8         & 13.6         & 32.0          & 22.4          & 30.4          & 22.4          \\ \cmidrule(l){2-12} 
                          & DGIR              & \textbf{34.4} & \textbf{26.4} & \textbf{34.4} & \textbf{27.2} & \textbf{42.4} & \textbf{28.0} & \textbf{44.5} & \textbf{27.7}  & \textbf{51.2} & \textbf{36.8} \\ 
                          \midrule
\multirow{7}{*}{IMCBench} & Base                    & 50.0          & 5.0           & 50.0          & 5.0          & 50.0         & 5.0          & 55.0          & 10.0          & 50.0          & 5.0           \\
                          & RAG (k=1)               & 60.0          & 25.0          & 55.0          & 20.0         & 50.0         & 5.0          & 60.0          & 25.0          & 50.0          & 5.0           \\
                          & RAG (k=3)               & 55.0          & 10.0          & 55.0          & 20.0         & 55.0         & 5.0          & 60.0          & 25.0          & 50.0          & 10.0          \\
                          & RAG (k=5)               & 60.0          & 25.0          & 60.0          & 35.0         & 55.0         & 10.0         & 60.0          & 40.0          & 55.0          & 20.0          \\
                          & File Context            & 85.0          & 40.0          & 70.0          & 35.0         & \textbf{80.0} & \textbf{45.0} & 80.0          & 55.0          & 85.0          & 50.0          \\ \cmidrule(l){2-12} 
                          & DGIR              & \textbf{95.0} & \textbf{65.0} & \textbf{85.0} & \textbf{45.0} & \textbf{80.0}  & \textbf{30.0} & \textbf{100.0} & \textbf{70.0} & \textbf{95.0} & \textbf{70.0} \\ 
                          \midrule
\multirow{7}{*}{TOTAL}    & Base                    & 10.3          & 2.8           & 13.8          & 4.8          & 14.5         & 4.8          & 15.9          & 6.2           & 14.5          & 4.8           \\
                          & RAG (k=1)               & 13.8          & 6.2           & 15.2          & 8.3          & 13.8         & 3.4          & 18.6          & 8.3           & 15.9          & 4.8           \\
                          & RAG (k=3)               & 16.6          & 5.5           & 14.5          & 7.6          & 15.2         & 3.4          & 18.6          & 8.3           & 15.9          & 5.5           \\
                          & RAG (k=5)               & 14.5          & 6.2           & 14.5          & 8.3          & 14.5         & 5.5          & 21.4          & 12.4          & 17.9          & 8.3           \\
                          & File Context            & 30.3          & 18.6          & 25.5          & 18.6         & 29.0         & \textbf{17.9} & 38.6          & 26.9          & 37.9          & 26.2          \\ \cmidrule(l){2-12} 
                          & DGIR              & \textbf{42.8} & \textbf{31.7} & \textbf{41.4} & \textbf{29.7} & \textbf{47.6} & \textbf{28.3} & \textbf{52.2} & \textbf{33.6}  & \textbf{57.2} & \textbf{41.4} \\ 
                          \bottomrule
\end{tabular}
\renewcommand{\arraystretch}{1}
\vspace{-0.3cm}
\label{tab:rq1}
\end{table*}

\section{Evaluation Result}

\subsection{RQ1: Effectiveness Evaluation}

\subsubsection{Effectiveness of DGIR in C-to-Rust Translation}

We evaluate the effectiveness of DGIR on the  C-to-Rust translation task using two benchmark datasets: DCBench and IMCBench. These datasets are derived from real-world open-source projects, representing diverse levels of complexity. The detailed results are presented in~\autoref{tab:rq1}. \gwj{ Based on the experimental results, we make the following key observations:}

\gwj{\textbf{(1)DGIR achieves superior overall performance by effectively resolving repository-level dependencies. } On the structurally complex DCBench, }
DGIR consistently outperforms all baselines across all model sizes. For instance, with Qwen2.5-Coder-32B, DGIR achieves a compilation success rate of 51.2\% and a computational accuracy of 36.8\%, outperforming the best-performing baseline File Context by 20.8\% and 14.4\%, respectively.  \gwj{This gap underscores that merely providing file context is insufficient for capturing the global dependency information necessary for repository-level consistency. } 
On IMCBench, 
\gwj{while reduced complexity benefits most methods, DGIR remains superior in functional correctness.} With Qwen2.5-Coder-14B and 32B, DGIR achieves a computational accuracy of 70.0\%, exceeding the best baseline by up to 20\%. Even with DSC-6.7B, the computational accuracy improves from 40.0\% (File Context) to 65.0\% with DGIR. \gwj{These results underscore that by modeling global project structures and integrating feedback-driven repair, DGIR not only facilitates syntactic correctness but also enhances the preservation of semantic equivalence.}

\gwj{\textbf{(2) DGIR exhibits strong generalization across varying model capacities. }}
For example, with DSC-6.7B, the compilation success rate increases from 30.3\% (File Context) to 42.8\%, and the computational accuracy from 18.6\% to 31.7\%. Similar trends are observed for larger models: under Qwen2.5-Coder-32B, the compilation success rate reaches 57.2\% and the computational accuracy 41.4\%, showing absolute improvements of 19.3\% and 15.2\% over File Context, respectively. These results confirm that DGIR not only benefits from large models but also substantially boosts the performance of smaller, more resource-efficient models.


\finding{1}{
\gwj{DGIR improves translation quality across diverse benchmarks and model scales, achieving absolute gains of up to 15.2\% in CA and 19.3\% in CSR compared to the strongest baselines. The framework demonstrates strong generalization, effectively empowering both large-scale and resource-efficient models.}}

\subsubsection{Effectiveness of \train in C-to-Rust translation}

As shown in \autoref{tab:model_compare}, \train demonstrates the effectiveness of our two-stage training methodology across both datasets. \gwj{Based on the results, we make the following observations:}

\gwj{\textbf{(1) \train achieves state-of-the-art overall performance across all evaluated metrics.}  In total, \train delivers the best aggregate results with 60.7\% CSR, 43.5\% CA, and a 65.4 CodeBLEU score, outperforming all baseline models. On IMCBench, it achieves a 90.0\% CSR and the highest CodeBLEU of 62.2, showcasing its superior ability to maintain semantic equivalence during translation. Compared to its base model QWC-7B, \train improves CA from 30.0\% to 60.0\%, effectively doubling the functional correctness.}

\gwj{\textbf{(2) \train outperforms baseline models with  larger parameter scales. }  For example, on the DCBench, \train achieves a 56.0\% CSR and 40.8\% CA, notably outperforming the much larger QWC-32B (51.2\% CSR, 36.8\% CA). This confirms that our RAST paradigm allows smaller models to capture intricate repository-level logic more effectively than general-purpose larger models.}




\finding{2}{\train markedly outperforms baseline models across all metrics, achieving comparable performance to much larger models through our two-stage training approach, demonstrating the practical effectiveness of integrating multi-task learning with compiler feedback.}

\subsection{RQ2: Ablation Studies}
\begin{table*}[t]
\centering
\caption{Evaluation results of \train compared with baseline models on the DCBench and IMCBench datasets.
}
\vspace{-0.3cm}
\renewcommand{\arraystretch}{0.85}
\begin{tabular}{@{}c|ccc|ccc|ccc@{}}
\toprule
\multirow{2}{*}{Model} & \multicolumn{3}{c|}{DCBench}                  & \multicolumn{3}{c|}{IMCBench}                  & \multicolumn{3}{c}{TOTAL}                     \\ \cmidrule(l){2-10} 
                       & CSR           & CA            & CodeBLEU      & CSR            & CA            & CodeBLEU      & CSR           & CA            & CodeBLEU      \\ \midrule
DSC-6.7B               & 34.4          & 26.4          & 55.8          & 95.0           & 65.0          & 54.4          & 42.8          & 31.7          & 55.6          \\
DSC-33B                & 34.4          & 27.2          & 60.1          & 85.0           & 45.0          & 57.1          & 41.4          & 29.7          & 59.7          \\
QWC-7B                 & 42.4          & 28.0          & 59.0          & 80.0           & 30.0          & 52.7          & 47.6          & 28.3          & 58.2          \\
QWC-14B                & 44.5          & 27.7          & 61.2          & \textbf{100.0} & \textbf{70.0} & 56.5          & 52.2          & 33.6          & 60.6          \\
QWC-32B                & 51.2          & 36.8          & 61.8          & 95.0           & \textbf{70.0} & 58.9          & 57.2          & 41.4          & 61.4          \\ \midrule
\train        & \textbf{56.0} & \textbf{40.8} & \textbf{65.9} & 90.0           & 60.0          & \textbf{62.2} & \textbf{60.7} & \textbf{43.5} & \textbf{65.4} \\ \bottomrule
\end{tabular}
\renewcommand{\arraystretch}{1}
\vspace{-0.3cm}
\label{tab:model_compare}
\end{table*}
\subsubsection{Ablation Study of DGIR}
\begin{table*}[!t]
\centering
\caption{Ablation study on DGIR. 
}
\vspace{-0.3cm}
\renewcommand{\arraystretch}{0.9}
\begin{tabular}{@{}c|c|cc|cc|cc|cc|cc@{}}
\toprule
\multirow{2}{*}{Dataset}  & \multirow{2}{*}{Method} & \multicolumn{2}{c|}{DSC-6.7B} & \multicolumn{2}{c|}{DSC-33B}  & \multicolumn{2}{c|}{QWC-7B}   & \multicolumn{2}{c|}{QWC-14B}   & \multicolumn{2}{c}{QWC-32B}   \\ \cmidrule(l){3-12} 
                          &                         & CSR       & CA          & CSR       & CA           & CSR       & CA           & CSR        & CA           & CSR       & CA           \\ \midrule
\multirow{5}{*}{DCBench}  & Plain Deps              & 22.4          & 19.2          & 24.8          & 20.2          & 23.2          & 15.2          & 40.7           & 30.1          & 42.4          & 33.6          \\
                          & w/o Compile             & 28.8          & 20.0          & 22.4          & 19.2          & 36.8          & 26.4          & 34.4           & 26.4          & 36.0          & 28.8          \\
                          & w/o Consistency         & \textbf{34.4} & 23.2          & \textbf{34.4} & \textbf{28.0} & 39.2          & \textbf{28.0} & \textbf{45.5}  & 33.3          & 44.0          & 35.2          \\
                          & w/o Repair              & 20.8          & 17.6          & 28.8          & 27.2          & 29.6          & 23.2          & 42.4           & \textbf{33.6} & 36.8          & 28.8          \\ \cmidrule(l){2-12} 
                          & DGIR               & \textbf{34.4} & \textbf{26.4} & \textbf{34.4} & 27.2          & \textbf{42.4} & \textbf{28.0} & 44.5           & 27.7          & \textbf{51.2} & \textbf{36.8} \\ \midrule
\multirow{5}{*}{IMCBench} & Plain Deps              & \textbf{95.0} & 55.0          & \textbf{90.0} & \textbf{60.0} & 70.0          & \textbf{35.0} & 85.0           & 45.0          & 85.0          & 50.0          \\
                          & w/o Compile             & 80.0          & 50.0          & 70.0          & 45.0          & 65.0          & 30.0          & 70.0           & 35.0          & 80.0          & 45.0          \\
                          & w/o Consistency         & \textbf{95.0} & 60.0          & 85.0          & \textbf{60.0} & 65.0          & 30.0          & 80.0           & 40.0          & 80.0          & 45.0          \\
                          & w/o Repair              & 75.0          & 50.0          & 70.0          & 40.0          & 75.0          & 25.0          & 85.0           & 55.0          & 80.0          & 65.0          \\ \cmidrule(l){2-12} 
                          & DGIR               & \textbf{95.0} & \textbf{65.0} & 85.0          & 45.0          & \textbf{80.0} & 30.0          & \textbf{100.0} & \textbf{70.0} & \textbf{95.0} & \textbf{70.0} \\ \midrule
\multirow{5}{*}{TOTAL}    & Plain Deps              & 32.4          & 24.1          & 33.8          & 25.7          & 29.7          & 17.9          & 46.8           & 32.1          & 48.3          & 35.9          \\
                          & w/o Compile             & 35.9          & 24.1          & 29.0          & 22.8          & 40.7          & 26.9          & 39.3           & 27.6          & 42.1          & 31.0          \\
                          & w/o Consistency         & \textbf{42.8} & 28.3          & \textbf{41.4} & \textbf{32.4} & 42.8          & \textbf{28.3} & 50.3           & 34.3          & 49.0          & 36.6          \\
                          & w/o Repair              & 28.3          & 22.1          & 34.5          & 29.0          & 35.9          & 23.4          & 48.3           & \textbf{36.6} & 42.8          & 33.8          \\ \cmidrule(l){2-12} 
                          & DGIR               & \textbf{42.8} & \textbf{31.7} & \textbf{41.4} & 29.7          & \textbf{47.6} & \textbf{28.3} & \textbf{52.2}  & 33.6          & \textbf{57.2} & \textbf{41.4} \\ \bottomrule
\end{tabular}
\renewcommand{\arraystretch}{1}
\vspace{-0.3cm}
\label{tab:rq2}
\end{table*}


To evaluate the contribution of each component in DGIR, we design four ablated variants: \textit{Plain Deps} (removing structural rewriting and import recovery), \textit{w/o Compile} (removing compiler feedback), \textit{w/o Consistency} (removing semantic consistency checking), and \textit{w/o Repair} (removing the iterative repair loop). Experimental results are presented in~\autoref{tab:rq2}.

\textbf{Experimental results confirm that every component of DGIR is indispensable. }
First, removing structural rewriting (\textit{Plain Deps}) causes a clear decline in overall accuracy; for instance, the Total CA on QWC-32B drops from 41.4\% to 35.9\%, indicating that fine-grained structural alignment is fundamental for capturing effective repository context. 
Second, removing compiler feedback (\textit{w/o Compile}) directly impacts syntactic validity; specifically, the compilation success rate (CSR) on DCBench with QWC-14B decreases  from 44.5\% to 34.4\%, proving that compiler-driven repair is essential for
high compilation rates. 
Third, excluding semantic consistency checks (\textit{w/o Consistency}) significantly compromises functional correctness. Notably, the functional accuracy on IMCBench (QWC-32B) drops sharply from 70.0\% to 45.0\%. Interestingly, this performance decay is negligible on smaller-scale models, suggesting that consistency guidance is more effective for models with stronger reasoning capabilities, whereas it may be considered optional for smaller models with limited self-correction potential.
Finally, removing the entire iterative repair loop (\textit{w/o Repair}) results in the most severe degradation across metrics; for example, the CSR on DCBench (QWC-7B) plummets from 42.4\% to 29.6\%, validating that the iterative refinement process is the core engine for bridging the gap between initial translation and executable, correct code.


\subsubsection{Ablation Study of \train}
\label{sec:ablation_train}

We further analyze the impact of different reward weighting strategies in the reinforcement learning phase. As shown in \autoref{tab:train_ablation}, we evaluate four \train variants with distinct reward configurations.

\textbf{A balanced reward configuration is critical for achieving robust translation.}
Specifically, the \textit{Compilation-Only (1:0)} strategy causes performance degradation, evidenced by a near-perfect Total CSR (97.9\%) but a complete failure in functional correctness (0.0\% Total CA), as the model learns to exploit the reward mechanism by generating trivial, safe outputs (e.g., \texttt{panic!}) to satisfy compilation constraints without implementing actual logic. 
Conversely, exclusively prioritizing or over-emphasizing alignment rewards (\textit{Test-Only 0:1} and \textit{Test-Heavy 1:2}) compromises syntactic reliability; for instance, while the 1:2 variant achieves the highest semantic similarity (Total CodeBLEU 66.5\%), it suffers a decline in computational accuracy compared to the balanced setting (40.8\% vs. 43.5\%). 
Ultimately, the \textit{Balanced (1:1)} configuration demonstrates superior robustness, achieving the highest Total CA of 43.5\% alongside a competitive compilation rate, confirming that the joint optimization of syntactic validity and functional alignment is requisite for high-fidelity migration.

\finding{3}{The ablation studies indicate that both the modular components of DGIR and the balanced reward configuration in \train are important. Structural guidance and iterative repair contribute complementary benefits, while reward balancing between compilation and test execution helps mitigate degenerate behaviors and improves translation quality.}

\subsection{RQ3: Industrial Project-level Migration}
\begin{table*}[t]
\centering
\caption{Ablation study on different reward weighting strategies during training. All models are based on the \train variant. \train-$\alpha$:$\beta$ denotes a model trained with compilation and test rewards weighted by $\alpha$ and $\beta$, respectively.
}
\vspace{-0.3cm}
\renewcommand{\arraystretch}{0.9}
\begin{tabular}{@{}l|ccc|ccc|ccc@{}}
\toprule
\multicolumn{1}{c|}{\multirow{2}{*}{Model}} & \multicolumn{3}{c|}{DCBench}               & \multicolumn{3}{c|}{IMCBench}               & \multicolumn{3}{c}{TOTAL}                  \\ \cmidrule(l){2-10} 
\multicolumn{1}{c|}{}                       & CSR        & CA          & CodeBLEU         & CSR        & CA          & CodeBLEU         & CSR        & CA          & CodeBLEU         \\ \midrule
\train-1:0                          & \textbf{97.6} & 0.0         & 47.2              & \textbf{100.0} & 0.0         & 37.8              & \textbf{97.9} & 0.0         & 45.9              \\
\train-0:1                          & 53.6       & 38.2        & 66.0              & 90.0       & 50.0        & \textbf{64.6}     & 58.6       & 39.8        & 65.8              \\
\train-1:1                          & 56.0       & \textbf{40.8} & 65.9              & 90.0       & \textbf{60.0} & 62.2              & 60.7       & \textbf{43.5} & 65.4              \\
\train-1:2                          & 56.1       & 39.3        & \textbf{66.8}     & 90.0       & 50.0        & 64.4              & 60.8       & 40.8        & \textbf{66.5}     \\ \bottomrule
\end{tabular}
\renewcommand{\arraystretch}{1}
\label{tab:train_ablation}
\vspace{-0.3cm}
\end{table*}

\begin{table}[]
\caption{
HWBench evaluation against baselines.
}
\vspace{-0.3cm}
\resizebox{\linewidth}{!}{%
\begin{tabular}{@{}c|ccccc|c@{}}
\toprule
Method   & DSC-6.7B & DSC-33B & QWC-7B & QWC-14B & QWC-32B & \train \\ \midrule
Base     & 0        & 0       & 0      & 0       & 1       & -           \\
Sactor   & 1        & \textbf{3}       & 2      & 2       & 5       & -           \\ \midrule
DGIR & \textbf{2}        & \textbf{3}       & \textbf{3}      & \textbf{6}       & \textbf{7}       & \textbf{7}           \\ \bottomrule
\end{tabular}
}
\vspace{-0.4cm}
\label{tab:rq4}
\end{table}

To evaluate \method in industrial contexts, we introduce \textsc{HWBench}, a dataset of 15 Huawei C projects with code sizes ranging from 1,000 to 3,000 lines. The evaluation metric is the number of projects that can be successfully built (\textit{Build}), reflecting practical applicability in real-world enterprise settings. The results are summarized in ~\autoref{tab:rq4}.

\gwj{\textbf{\method demonstrates superior robustness and practical effectiveness in industrial-scale migration. }} 
On HWBench, the \textit{Base} strategy nearly fails across all configurations, while \textit{Sactor} achieves only modest improvements (e.g., 5 builds on QWC-32B). In contrast, DGIR independently attains superior performance, with 6 and 7 successful builds on QWC-14B and QWC-32B, respectively, even without RAST adaptation. When RAST is integrated, \train's successful builds increase from 3 to 7, effectively matching the performance of the much larger QWC-32B model. This underscores both the robustness of our inference framework and the efficiency of our task-specific tuning.

\finding{4}{\method demonstrates strong robustness and practical effectiveness in industrial-scale C-to-Rust migration, notably improving build success rates and supporting enterprise-level code translation.}

\section{Discussion}

\begin{figure}[]
    \centering
    \includegraphics[width=\linewidth]{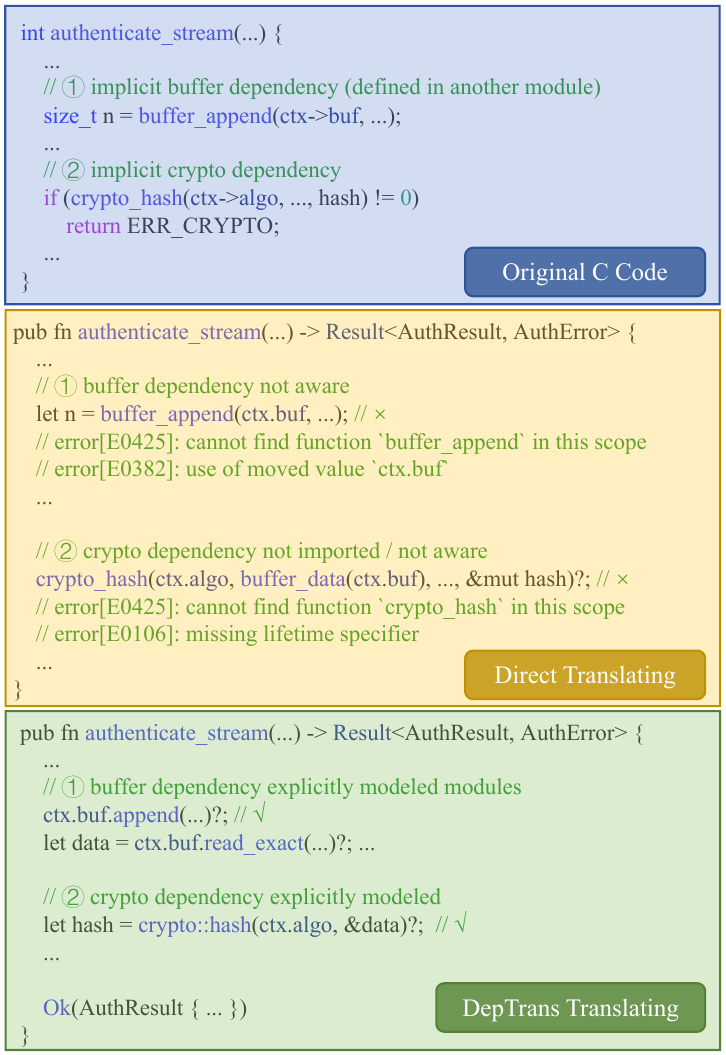}
    \caption{Case study of authenticate\_stream translation generated by Qwen2.5-Coder-7B.
    }
    \vspace{-0.3cm}
    \label{fig:case}
\end{figure}

\subsection{Case Study}

We present a case study on the \texttt{authenticate\_stream} function to illustrate the effectiveness of \method (\autoref{fig:case}). This function relies on implicit buffer and cryptographic dependencies defined in external modules, which pose a significant challenge for translation approaches lacking project-wide context. As shown in \autoref{fig:case}, baseline methods exhibit critical failures: direct translation produces scope errors (e.g., \texttt{error[E0425]}) and ownership violations such as \texttt{use of moved value}, as it fails to recognize external module interfaces and Rust's strict borrowing rules. In contrast, \method generates a syntactically valid and idiomatic translation. By explicitly modeling cross-module dependencies and mapping them to a semantically aligned Rust dependency pool, \method correctly resolves module-qualified calls (e.g., \texttt{crypto::hash}) and ensures functional integrity. This demonstrates \method's superior ability to navigate the structural complexities of repository-scale C-to-Rust migration.

\subsection{\wcz{Impact of Repair Iterations} in DGIR}
\begin{figure}[]
    \centering
    \includegraphics[width=\linewidth]{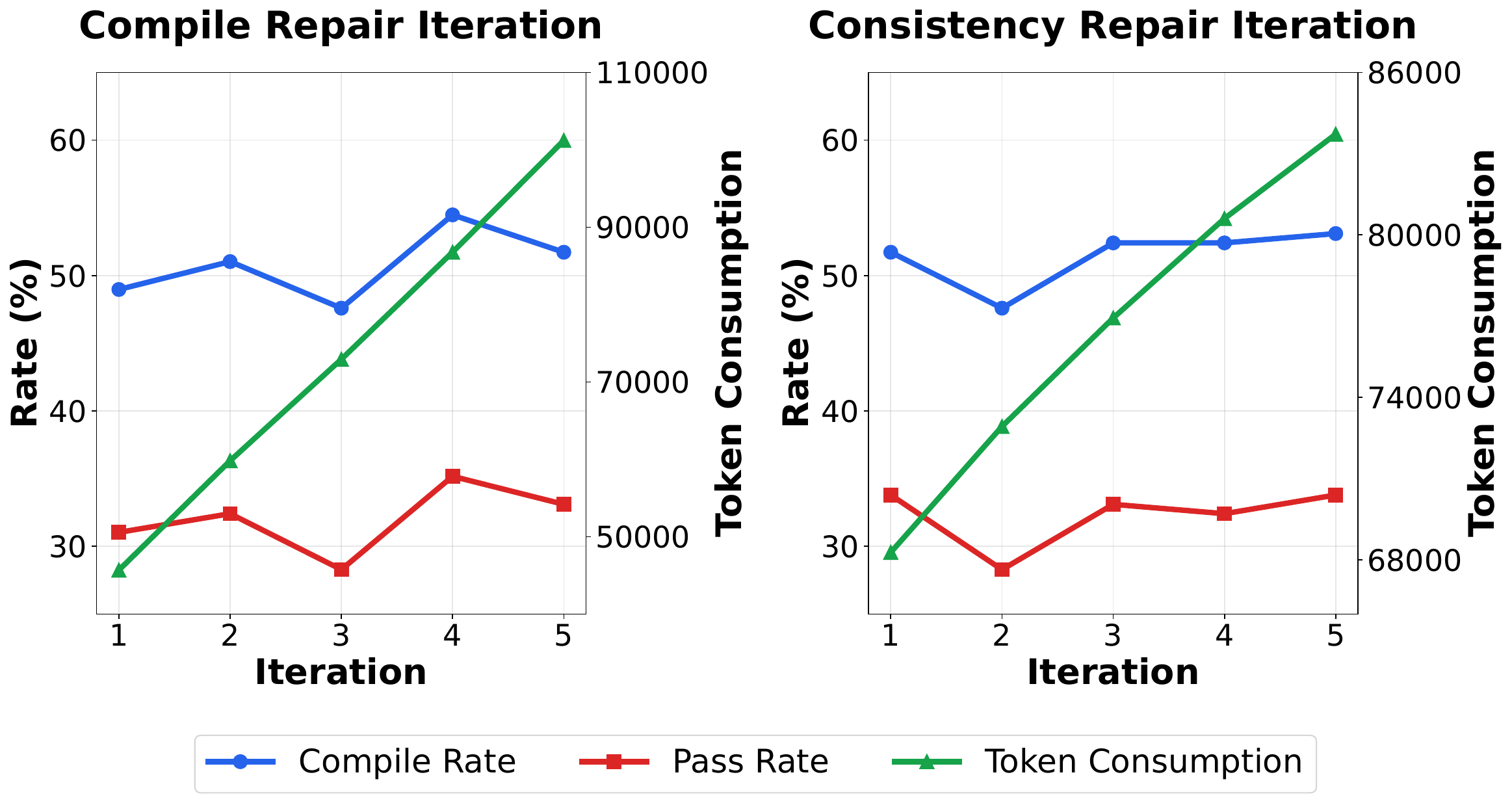}
    \vspace{-0.55cm}
    \caption{Impact of compilation repair and consistency repair iterations on translation performance and average token consumption.}
    \label{fig:iteration}
     \vspace{-0.3cm}
\end{figure}

We evaluate the impact of repair iteration counts on Qwen2.5-Coder-7B by varying compiler repair (1--5) with fixed consistency repair (2), and varying consistency repair (1--5) with fixed compiler repair (3). As shown in~\autoref{fig:iteration}, we report the compilation rate, accuracy, and token consumption for each configuration.

\textbf{Increasing repair iterations improves performance but incurs higher costs. } For compiler-guided repair, computational accuracy rises from 31.03\% at iteration 1 to a peak of 35.17\% at iteration 4, yet this comes with a sharp increase in token consumption exceeding 100k at iteration 5. Similarly, consistency-guided repair shows marginal improvements, where iteration 5 achieves the highest compilation rate of 53.10\% but consumes more tokens than iteration 1 which attains identical computational accuracy.


To balance translation performance with inference overhead, we recommend setting the compiler-guided repair iteration count to 2--3 and the consistency repair iteration count to 1 or 3. This configuration maximizes computational efficiency while maintaining robust migration quality in real-world pipelines.

\subsection{Effectiveness on Open-Source Benchmarks}
\begin{table}[]
\caption{
CRust-Bench evaluation against baselines.
``Build'' indicates the number of successfully compiled projects, and ``Test'' represents the number of projects passing all test cases. 
}
\vspace{-0.3cm}
\resizebox{\linewidth}{!}{%
\renewcommand{\arraystretch}{0.85}
\begin{tabular}{@{}c|ccc|ccc@{}}
\toprule
\multirow{2}{*}{Model} & \multicolumn{3}{c|}{Build} & \multicolumn{3}{c}{Test} \\ \cmidrule(l){2-7} 
                       & Base & Sactor & DGIR & Base & Sactor & DGIR \\ \midrule
DSC-6.7B               & 0    & 4      & \textbf{6}                & 0    & \textbf{0} & \textbf{0}                \\
DSC-33B                & 1    & \textbf{12} & 10                        & 0    & \textbf{2} & 1                         \\
QWC-7B                 & 0    & 5      & \textbf{11}               & 0    & \textbf{2} & \textbf{2}                \\
QWC-14B                & 1    & 8      & \textbf{20}               & 0    & 2          & \textbf{4}                \\
QWC-32B                & 6    & 12     & \textbf{26}               & 0    & 4          & \textbf{5}                \\ \bottomrule
\train         & -    & -           & \textbf{27} & -    & -          & \textbf{5} \\ \bottomrule
\end{tabular}%
\renewcommand{\arraystretch}{1}
}
\vspace{-0.3cm}
\label{tab:rq3}
\end{table}


We evaluate \method on CRust-Bench~\cite{CRUST-Bench}, which comprises 100 open-source C projects with manually written Rust test cases.

\textbf{\method demonstrates superior generalization and robustness on open-source benchmarks. }As shown in Table~\ref{tab:rq3}, \method consistently outperforms baselines across all metrics. Specifically, with QWC-32B, it compiles 26 projects, doubling the 12 builds by Sactor, and increases test-passing projects to 5. Integrating \train yields state-of-the-art results with 27 builds and 5 passes, confirming the system's robustness on complex repositories.

\subsection{How safe is the Rust code generated by \method?}
\begin{table}[]
\caption{Evaluation of unsafe code ratio (\%) in translated Rust}
\vspace{-0.3cm}
\renewcommand{\arraystretch}{0.9}
\begin{tabular}{@{}c|cccc@{}}
\toprule
Model       & DCBench & IMCBench & CRustBench & HWBench \\ \midrule
DSC-6.7B    & 0.98    & 0        & 3.39       & 2.58    \\
DSC-13B     & 0.55    & 0        & 0.1        & 1.72    \\
QWC-7B      & 1.015   & 0        & 1.26       & 2.35    \\
QWC-14B     & 0.44    & 0        & 0.24       & 1.46    \\
QWC-32B     & 0       & 0        & 0.25       & 0.76    \\ \midrule
\train & 0.32    & 0        & 0.54       & 1.11    \\ \bottomrule
\end{tabular}
\renewcommand{\arraystretch}{1}
\vspace{-0.3cm}
\label{tab:security}
\end{table}

To assess the safety of Rust code produced by \method, we measure the proportion of ``unsafe'' lines in translated projects, as summarized in ~\autoref{tab:security}. Lower values indicate safer code, i.e., fewer instances requiring explicit ``unsafe'' blocks.

\textbf{\method consistently minimizes unsafe code ratios across all benchmarks.} For instance, the unsafe code ratio remains below 3\% for all models on DCBench and HWBench. Even on the complex CRustBench, the ratio stays remarkably low, exemplified by 3.39\% for DSC-6.7B, which confirms the framework's capability to effectively reduce memory safety risks and ensure maintainability.



\subsection{Threats To Validity}

\textbf{Limited LCMs.} We reduce selection bias by evaluating five diverse LCMs with different sizes. Moreover, our core contributions, including dependency enhancement and iterative repair, are model-agnostic and can be easily applied to other architectures.

\textbf{Potential Data Leakage.} Since pretraining data is not public, we cannot fully rule out leakage. However, the poor performance of simple prompting suggests this is not a major issue. The significant gains from our framework confirm that the improvements come from our method rather than the model just memorizing the data.




\section{Related Work}

Existing C-to-Rust migration approaches follow two primary paradigms: rule-based and LLM-based. Rule-based approaches, such as C2Rust~\cite{c2rust_website}, leverage compiler infrastructures to ensure structural equivalence, while subsequent works incorporate static analysis~\cite{concrat,ownership} and alias emulation~\cite{Emre2021,Emre2023} to enhance safety. Although these tools provide high functional fidelity, the resulting code is often unidiomatic and fails to fully utilize Rust's safety guarantees.

LLM-based approaches aim to bridge this gap by learning idiomatic patterns from large-scale corpora. Recent studies have explored direct translation with feedback loops~\cite{VERT,Syzygy,DBLP:journals/corr/abs-2503-12511} or hybrid refinement strategies that post-process rule-based outputs~\cite{DBLP:journals/corr/abs-2501-14257,LAC2R}. These diverse explorations have collectively highlighted two persistent, open challenges in the field: the difficulty of maintaining coherence across complex repository-level dependencies and the scarcity of high-quality parallel data. Our work, \method, specifically addresses these architectural and data-level bottlenecks.

\section{Conclusion}

In this paper, We propose a novel approach for repository-level C-to-Rust translation. It consists of two key components: a dependency-guided translation framework and a two-stage training paradigm. The framework extracts fine-grained dependencies from C code, maps them to their Rust equivalents, and constructs precise translation context to guide LLM-based generation. It further applies compiler diagnostics and LLM-based consistency checks for iterative error repair. The training paradigm combines multi-task fine-tuning with compiler-feedback reinforcement learning, supported by our constructed dataset of function-level and repository-level C-Rust pairs. Extensive experiments show that our approach consistently improves compilation success and functional correctness over strong baselines, demonstrating its effectiveness for practical migration scenarios.

\section*{Acknowledgment}
This research is supported by National Natural Science Foundation of China under project (No. 62472126) and CCF-Huawei Populus Grove Fund.

\bibliographystyle{ACM-Reference-Format}
\bibliography{acmart}

\end{document}